\documentclass[aps,prl,twocolumn,groupedaddress,showpacs]{revtex4}

\usepackage[latin1]{inputenc}
\usepackage[english]{babel}
\usepackage{setspace}  
\usepackage{graphicx}
\usepackage{amssymb}
\usepackage{amsmath}
\usepackage{bbold}
\usepackage{amscd}
\usepackage{rotating}
\usepackage{color}
\usepackage{epstopdf}

\newcommand{\ket}[1]{\left\vert{#1}\right\rangle}

\newcommand{\abs}[1]{\left\vert #1 \right\vert}

\newcommand{\erf}[1]{Eq.~(\ref{#1})}

\newcommand{\Heff}{H_{\rm eff}}

\newcommand{\order}[1]{\mathcal{O}\left( #1 \right)}

\newcommand{\T}{\mathsf{T}}
\newcommand{\M}{\mathsf{M}}

\newcommand{\prsec}[1]{\noindent {\emph{#1}}.-- }
\begin{document}
\title{Stroboscopic Generation of Topological Protection}

\author{C. M. Herdman$^1$, Kevin C. Young$^1$}
\affiliation{
Berkeley Center for Quantum Information and Computation, Departments of Physics$^{1}$ and Chemistry$^{2}$, University of California, Berkeley, California 94720, USA\\
$^3$Theoretische Physik, ETH Zurich, 8093 Zurich, Switzerland
}
\author{V.W. Scarola$^{2,3}$}
\altaffiliation [Present Address: ] {Department of Physics, Virginia Tech, Blacksburg, Virginia 24061, USA}
\affiliation{
Berkeley Center for Quantum Information and Computation, Departments of Physics$^{1}$ and Chemistry$^{2}$, University of California, Berkeley, California 94720, USA\\
$^3$Theoretische Physik, ETH Zurich, 8093 Zurich, Switzerland
}
\author{Mohan Sarovar$^2$}
\affiliation{
Berkeley Center for Quantum Information and Computation, Departments of Physics$^{1}$ and Chemistry$^{2}$, University of California, Berkeley, California 94720, USA\\
$^3$Theoretische Physik, ETH Zurich, 8093 Zurich, Switzerland
}
\author{K. B. Whaley$^2$}
\affiliation{
Berkeley Center for Quantum Information and Computation, Departments of Physics$^{1}$ and Chemistry$^{2}$, University of California, Berkeley, California 94720, USA\\
$^3$Theoretische Physik, ETH Zurich, 8093 Zurich, Switzerland
}

\date{\today}

\begin{abstract}
Trapped neutral atoms 
offer a powerful route to robust simulation of complex quantum systems.  We present here a stroboscopic scheme for realization of a Hamiltonian with $n$-body interactions on a set of neutral atoms trapped in an addressable optical lattice, using only 1- and 2-body physical operations together with a dissipative mechanism that allows thermalization to finite temperature or cooling to the ground state. 
We demonstrate this scheme with application to the toric code Hamiltonian,  ground states of which can be used to robustly store quantum information when coupled to a low temperature reservoir.
\end{abstract}

\pacs{03.67.Pp, 03.65.Vf,03.67.Lx}

\maketitle

\prsec{Introduction}  
Among the most exciting aspects of quantum simulation is the possibility of generating and studying exotic quantum phases such as those possessing topological order that can be used to robustly store and process quantum information.
 The Hamiltonians  governing these phases frequently require more-than-2-body interactions that are hard or even impossible to realize naturally.  This difficulty has spurred much theoretical and experimental effort in the artificial engineering of Hamiltonians, particularly for trapped neutral atoms~\cite{lewenstein2007}.  Many proposals have been made for the generation of 2-body Hamiltonians using static emulation schemes and some experimental realizations have appeared ~\cite{greiner2002,spielman2008}. Specific proposals have appeared for generating $n$-body interactions ~\cite{buchler2007}, but have focused on static emulation.

 We present here an alternative, dynamic emulation approach to systematic generation of $n$-body interactions that is based on  sequences of control pulses
 which individually realize $1$- and $2$-body operations on internal atomic levels.
We show that this stroboscopic realization of the Hamiltonian can be implemented simultaneously with a dissipative thermalization protocol to stabilize the system from the effects of
imperfect quantum operations and environmental noise. In the zero temperature limit, this can be view as replacing algorithmic error correction in an equivalent quantum circuit model with a dissipative procedure to remove errors~\cite{Lloyd1999}. The resource requirements for this thermalization protocol are different from those of algorithmic error correction, and may be more accessible to experiment in the foreseeable future. We illustrate the approach here with stroboscopic generation of the 4-body toric code Hamiltonian, which constitutes one of the simplest exactly solvable models with a ground state topological phase~\cite{kitaev2003}:  
\begin{equation}
H_0^{TC}  = -J_e \sum_v \prod_{j\in v} \sigma^z_j -J_m \sum_{p} \prod_{j\in p}  \sigma^x_j \label{H_TC},
\end{equation}
where $\sigma_j$ denotes a Pauli operator on the links of a square lattice and $v/p$ denote the vertex/plaquette of the lattice.  The ground state of this model possesses topological order, and therefore has anyonic quasiparticle excitations and, on a lattice with periodic boundary conditions, an emergent topological degeneracy. Quantum information can be encoded in this ground state degeneracy and manipulated with controlled creation and braiding of anyons~\cite{kitaev2003, Nayak2008}.
  In a finite sized system~\cite{Nussinov08,Castelnovo07}, the topological order of the ground state and gap to excited states protects against decoherence and loss of quantum information due to noise \emph{provided the system is coupled to a low temperature bath}.  Our analysis below will provide a scheme for generating both $H_0^{TC}$ and an effective low temperature bath, 
realizing the topological protection characteristic of the toric code.

\begin{figure}[t] 
   \centering
  \includegraphics[width=3in]{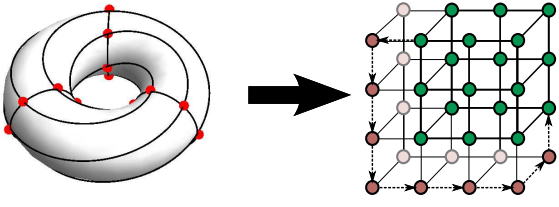} 
 \caption{ 
(Color online) An embedding of the toric code into a cubic lattice. Twisted-periodic boundary conditions are imposed by SWAP-gate shuttling along auxilliary sites, indicated by dashed-arrows.  Bold lines connect logical nearest-neighbors.~\cite{longpaper}
}
\label{fig2}
\end{figure}

The physical context for our analysis is a set of $\sim 250$ individual $^{133}$Cs atoms trapped at the sites of an addressable simple cubic optical lattice~\cite{nelson2007}. A lattice spacing of $5~\mu$m~\cite{nelson2007} allows essentially perfect addressability~\cite{beals2008}.
The orbital degrees of freedom are frozen on the time scales relevant to our analysis and we need consider only internal atomic degrees of freedom. Two hyperfine levels (e.g., $\ket{F,m_F} = \ket{4,4}, \ket{3,3} $) define a 2-level pseudospin system.  We realize $H_0^{TC}$ in the interaction representation defined by the pseudospin energies.
Auxiliary internal levels are used to realize 1-spin and 2-spin quantum operations, using optical frequency Raman pulses to generate arbitrary single-spin operations and excitation of one atom to a Rydberg state, e.g., the $n\approx 80$ state, to generate controlled-phase gates, CPHASE 
~\cite{jaksch2000}.  To achieve 
thermalization or 
cooling, the Hamiltonian $H_0^{TC}$ is supplemented by coupling the primary system spins to an ancillary set of pseudospins that will be 
dissipatively controlled to simulate a  thermal reservoir.
Since the pseudospins are localized at the sites of a cubic lattice, one can choose to either realize $H_0^{TC}$ on a single plane using a surface code ~\cite{bravyi1998} or in a three-dimensional cubic array with toroidal boundary conditions realized by SWAP 
operations (see Fig.~\ref{fig2}).

\prsec{Effective Hamiltonian Evolution} Given the
ability to perform both Rydberg-induced CPHASE gates between atoms in
neighboring sites and arbitrary 1-body rotations, 
$\exp{(- i \theta
\sigma_j)}$, on individual atoms, where $\theta$ is a variable phase
angle, sequences of these operations can be chosen to
generate effective \emph{$n$-body} interactions through high-order
terms in the Magnus expansion
\cite{klarsfeld1989}, allowing stroboscopic simulation of a broad
class of Hamiltonians. Consider the operator sequence, $U_n U_{n-1}\cdots U_2U_1$,  where
the $U_j$ are the 1- or 2-body gates described above.
Effective interactions are found through:
\begin{align*}
	 \Heff(t) \equiv	& \;\frac{i\hbar}{t} \ln\!\left(U_n U_{n-1}\cdots
U_2 U_1\right)\\
	 		=	& \sum_j \frac{i \hbar}{t} \ln U_j - \sum_{j<k} \frac{i \hbar}{2
t} \left[\, \ln U_j, \ln U_k\right] + \order{\abs{\abs{\ln U}}^3}.
\end{align*}

Consider now simulation of the 4-body interactions in $H_0^{\rm TC} $.
We use the notation $U_j(\phi) \equiv e^{-i \phi \Sigma_j}$ and
define $\Sigma_1 =  \sigma^z  \sigma^y\sigma^0 \sigma^0$,  $\Sigma_2 =
\sigma^0 \sigma^x  \sigma^y  \sigma^0 $, and $ \Sigma_3 = \sigma^0
\sigma^0 \sigma^x  \sigma^z$, where $\sigma^0$ is the identity
operator.  For simplicity, it is assumed that each $U_j(\alpha)$ takes
a time $\tau$ to execute.
We construct the operator sequence,
\begin{equation}
	 U_{123}(\alpha, \beta,\gamma) = U_{12}(\alpha,\beta)  U_3(\gamma)
U_{12}^\dagger(\alpha,\beta) U_3^\dagger(\gamma),
	\label{eq:sequence}
\end{equation}
where $U_{12}(\alpha,\beta) = U_2(\beta) U_1(\alpha)
U_2^\dagger(\beta) U_1^\dagger(\alpha) $.  This sequence
acts over a time $10 \tau$ to generate the following effective
Hamiltonian at a single vertex, $v$:
\begin{align*}
	 H_{\text{eff}}^{zzzz} =& J_e \prod_{j\in v} \sigma^z_j
	 +\frac{\chi}{ \alpha\gamma} \left(\alpha \left[\sigma^0 \sigma^x
\sigma^z\sigma^z \right]_{v}+ \gamma
\left[\sigma^z\sigma^z\sigma^y\sigma^0\right]_{v}\right)\\
	 &+\chi  \left( \left[\sigma^0 \sigma^x \sigma^y
\sigma^0\right]_{v}-2\beta/\gamma \left[\sigma^0 \sigma^0 \sigma^x
\sigma^z\right]_{v}
	 	\right)  +\order{\phi^6}
\end{align*}
where $[O]_{v/p}$ denotes the application of the (up to) four-body
operator $O$ to the spins meeting at a vertex $v$ or surrounding a
plaquette $p$; we choose $\abs{\alpha} = \abs{\beta} = \abs{\gamma}
\equiv \phi$; and $J_e = \chi \left(1 - 3\phi^{2}\right)/\alpha\gamma+
\mathcal{O}(\phi^6)$ with  $\chi\equiv\alpha^2\beta\gamma^2
\left(2\hbar/ 5\tau\right)$.  By repeating the operator sequence a
second time with sign reversals $\alpha\rightarrow-\alpha$ and
$\gamma\rightarrow-\gamma$, we cancel the 
fourth order terms in $\phi$,
giving $U_{123}(-\alpha,\beta,-\gamma)U_{123}(\alpha,\beta,\gamma)$ 
that acts for a time $20\tau$ to generate the effective Hamiltonian
\begin{eqnarray}
	 H_{\text{eff}}^{zzzz} = J_e \prod_{j\in v} \sigma^z_j
	 + \chi \left[\sigma^0 \sigma^x \sigma^y \sigma^0\right]_{v}  +\order{\phi^6}.
\label{Heffzzzz}
\end{eqnarray}	
The sequence $U_{123}(-\alpha,\beta,-\gamma)U_{123}(\alpha,\beta,\gamma)$
is specifically designed to cancel the lowest-order 
($\phi^4$) perturbation terms
without affecting the gap.  The remaining $\phi^5$ term is a 2-body
perturbation to $H_0^{TC}$.
Repeating this sequence with appropriate sign reversals will cancel
these higher order terms.  However, the ground state subspace of
$H_0^{TC}$ is robust to these remaining perturbations (see below).  A
shorter operator sequence
may then be preferable to reduce gate errors. The
plaquette operator, $\Heff^{xxxx}$, can be generated by
cyclic permutation of the Pauli operators in the above expressions for
$\Sigma_1, \Sigma_2$ and $\Sigma_3$.

Simulation of $H_0^\text{TC}$ then requires
application of the pulse sequence to all vertices and plaquettes
conforming to a two dimensional square lattice with periodic boundary
conditions.
Vertex and plaquette terms may be applied serially as: $\exp{(-i
\Heff^{xxxx}t/\hbar)}\exp{(-i \Heff^{zzzz} t/\hbar)} \approx \exp{-i
\left(\Heff^{xxxx} + \Heff^{zzzz}\right) t/\hbar} $.  Because only the
perturbation terms fail to commute, the truncation error in the above
expression occurs at orders larger than $\phi^7$.  For 18 pseudospins,
representing a $3\times 3$ system with toroidal boundary conditions, a
completely serial implementation yields a stroboscopic cycle time of
720$\mu$s
using the estimate $\tau \sim
500$ns~\cite{Weiss_private_communication} and the minimal count of one
CPHASE and four 1-spin gates to realize all $U_j(\alpha)
$~\cite{zhang2003}. This may be reduced by implementing some operators in
parallel.

\prsec{Simulated Thermalization}
The pseudospin subspace of the system defined by the internal states of the trapped atoms will interact with the external environment through the controlled quantum operations in the above pulse sequences and uncontrolled noisy interactions. Noise in the optical lattice system will not drive the simulation subspace to a state that is thermal under the simulated Hamiltonian~\cite{brown2007}. Additionally, noise in the above sequence of control gates will add entropy and effectively heat the system.
The entropy production $\Delta\mathrm{S}$ resulting from imperfect gate operation $\Delta\mathrm{S} \sim \mathrm{EPG}$, where $\mathrm{EPG}$ is the error per gate~\cite{longpaper}.  
Quantum circuit models are usually supplemented by error correction schemes to effectively remove entropy from the system.  We take a different approach here, constructing 
an effective system-reservoir interaction to control system entropy and relax the system to the ground state or a thermal state. 

To maintain the simulated system at a thermal steady state we add an interaction $H_{\mathrm{sr}}$ of the system pseudospins with a set of ancillary pseudospins. In the optical lattice system, these ancillary pseudospins, which may be a 2$^{nd}$ species of atom, are trapped in an offset, 
intercalated optical lattice, such that each ancillary atom is adjacent to a system atom. Consider a Hamiltonian with local $n$-body interactions  of the form:
$
H_0 = - \sum_{\nu} J_\nu \sum_{\mathcal{N}} h^\nu_{\mathcal{N}}
$
, where $h^\nu_{\mathcal{N}}$ is an $n$-body operator involving a neighborhood of pseudospins, $\mathcal{N}$,  including pseudospin $i$, with eigenvalues $\pm 1$, $\nu$ labels the type of interaction, and $J_\nu$ is a constant. Additionally we define the pseudospin flip operator $\Sigma^\nu_i$ such that $\Sigma^\nu_i \vert h^\nu_{\mathcal{N}} = \pm 1 \rangle = \vert h^\nu_{\mathcal{N}} = \mp 1 \rangle$ when $i\in \mathcal{N}$. 
When all $[h^\nu_{\mathcal{N}},h^\mu_{\mathcal{N}'}] = 0$, as is the case for Eq.~(\ref{H_TC}), 
we can define 
$E^\dagger_{i,\nu} = \frac{1}{4}\Sigma^\nu_i \left( \mathbb{1} + h^\nu_{\mathcal{N}} \right)  \left( \mathbb{1} + h^\nu_{\mathcal{N}'} \right), 
T_{i,\nu} = \frac{1}{4} \Sigma^\nu_i \left( \mathbb{1} - h^\nu_{\mathcal{N}} \right) \left( \mathbb{1} + h^\nu_{\mathcal{N}'} \right)
$
with $i = \mathcal{N} \cap\mathcal{N}'$. These areare $(2n-1)$-body interactions; $E^\dagger_{i,\nu}$ creates a pair of excitations about $i$ and $T_{i,\nu}$ translates an excitation about $i$. The energy gap for creation of a pair of excitations is $\Delta_\nu = 4 J_\nu$.

A route to guaranteeing the thermal equilibration of this system is for it to evolve under the Lindblad master equation
$\dot{\rho} = -i/\hbar \left [ H_{0}, \rho \right] +L\left[ \rho \right]$,
where $\rho$ is the density matrix and $L\left[ \rho \right]$ is the superoperator
$
L\left[ \rho \right]=\sum_{\omega} \left ( 2c_\omega^{\vphantom{dagger}} \rho  c^{\dagger}_\omega-c^{\dagger}_\omega c_\omega^{\vphantom{dagger}}\rho  -\rho c^{\dagger}_\omega c_\omega^{\vphantom{dagger}} \right )
$, 
with $\{c_{\omega}^{\vphantom{dagger}}, c^{\dagger}_{\omega}\}$ the Lindblad operators.
With $\{c_{i,\nu}\}$ given by
\begin{equation}
\left\{ \sqrt{\frac{1-p }{2} \lambda^* }E_{i,\nu} ,\sqrt{ \frac{p}{2}\lambda^* }E^\dagger_{i,\nu} , \sqrt{ \frac{\gamma^*}{4} }T_{i,\nu}, \sqrt{ \frac{\gamma^*}{4} }T^\dagger_{i,\nu} \right\},
\label{jump_ops_sys}
\end{equation}
\noindent
the Lindblad master equation describes equilibration with a bath of temperature $T=-\Delta/\ln{(p)}$.
The unique stationary state of the system is then the thermal state under $H_0$ with temperature $T$. $\lambda^*$ and $\gamma^*$ are relaxation rates, and their values dictate the thermalization time.  For simplicity we have set $\Delta_{\nu}=\Delta$.  
To generate evolution under such a master equation, we introduce a set of non-interacting ancillary pseudospins that independently undergo strong dissipation.  Each local neighborhood of the system interacts locally with a single thermal ancillary pseudospin $\T_{i,\nu}$ and a single maximally mixed ancillary pseudospin $\M_{i,\nu}$ for each $\nu$:
$H_{\text{sr}}= g\sum_{\nu, i}  \left( E_{i,\nu}^{\dagger}\sigma_{\T_{i,\nu}}^- +T_{i,\nu}^{\vphantom{dagger}}  \sigma_{\M_{i,\nu}}^- +h.c.\right)
\label{Hsysres}$
The master equation of the system and ancilla pseudospins combined is of the above Lindblad form with Lindblad operators:
\begin{equation}
\left\{ c_{i,\nu} \right\}=\Bigl\{ \sqrt{\frac{1-p}{2} \lambda}\sigma^-_{\T_{i,\nu}} ,\sqrt{\frac{p}{2} \lambda}\sigma_{\T_{i,\nu}}^+ , \sqrt{\frac{\gamma}{4}}\sigma_{\M_{i,\nu}}^+,\sqrt{\frac{\gamma}{4}}\sigma_{\M_{i,\nu}}^- \Bigr\}, \nonumber
\end{equation}
where $\lambda$ and $\gamma$ define the relaxation rates of the individual ancillary pseudospins.
With this choice of $H_{\mathrm{sr}}$ it can be shown \cite{longpaper} that for $g \ll \lambda$ the system pseudospins evolve under a renormalized Lindblad master equation with $c_{\omega}$ given by \erf{jump_ops_sys} and $\lambda^* =  4\left(g/\hbar \right)^2\lambda $, thus leaving the thermal state of $H_0$ as the unique stationary state.  

For $p=0$, the effective system-reservoir interaction cools the system towards the ground state, and the Lindblad operators can be reduced to the $n$-body terms $\{\sqrt{\lambda^*}(E_{i,\nu}+T_{i,\nu}),\sqrt{\lambda^*}(E_{i,\nu}+T^\dagger_{i,\nu})\}$~\cite{Diehl08,kraus2008}.  In this limit the ancillary pseudospins become an effective low temperature bath with a cooling rate $\Gamma_{\text{c}}\sim \lambda^*$ and heating rate determined by gate errors and any environmental noise. Competition between these rates leads to a minimum reachable temperature for the system $T_{\text{min}}\sim \Delta /\ln\left(\Gamma_{\text{c}}/\Gamma_{\text{e}}\right)$, 
where $\Gamma_{\text{e}}\sim\text{EPG}\times \Omega$, with
EPG and $\Omega$ the error rate and frequency of application of $U_j(\alpha)$, respectively.  

The Lindblad master equation, with operators given by \erf{jump_ops_sys}, generates a \emph{unitary} system-reservior interaction but \emph{nonunitary} reservoir relaxation.   Stroboscopic simulation of $H_{\mathrm{sr}}$ is performed in a manner analogous to the $H_0^{TC}$ simulation described above.  Phase angles are chosen in the 1- and 2-body gates to generate an effective static interaction strength $g$ over the time $t_{sr}$ between applications of $H_{sr}$, such that $gt_{sr}/\hbar < \pi/2$.  Nonunitary evolution of the reservoir is generated by encoding the reservoir as two levels of a $\Lambda$-system.  The pseudospin states are the ground state $\ket{0}$ and the meta-stable state $\ket{1}$.  
State $\ket{2}$ is chosen to have fast spontaneous emission to $\ket{0}$, with rate $\Gamma_{20}$.  This spontaneous emission is the decoherence mechanism required to generate the nonunitary Lindblad evolution.   The ancillary pseudospin levels can be placed in a thermal state via the following procedure: i) $\pi$-pulse on the $\ket1\rightarrow \ket2$ transition. ii) Wait for decay to ground state, $\ket{0}$.   iii) $\pi$-pulse on $\ket0\rightarrow \ket1$ transition.  iv) $\theta$-pulse on the $\ket1\rightarrow \ket2$ transition. v) Wait for decay, which now yields the final pseudospin state, $\rho = {\rm diag}\left\{ \sin^2(\theta), \cos^2(\theta) \right\}$, corresponding to an effective temperature $T_{\rm eff} =\Delta/\left(2 \ln\left(\cot \theta \right)\right)$. The above stroboscopic procedure generates 
$\lambda^* \approx g^2t_{sr}/\hbar^2$ in \erf{jump_ops_sys}.  
The procedure can be simplified in the limit of cooling towards zero temperature by eliminating steps iii-v, 
when it becomes similar to the optical pumping scheme employed in measurement of qubit states for trapped ions~\cite{wineland1980}. This thermalization procedure is then repeated and interleaved with the stroboscopic application of 
$H_0$.

\prsec{Thermalization of the Toric Code}
$H_0^{\mathrm{TC}}$ is in the local form of $H_0$, with two types of excitations, electric charges and magnetic vortices, ($\nu=\text{e},\text{m}$) that reside on vertices and plaquettes, respectively, of the square lattice. The excitation operators are defined with $h_v^e = \prod_{j \in v} \sigma_j^z$, $h_p^m = \prod_{j \in p} \sigma_j^x$, and $\Sigma^{e,m}_i = \sigma^{x,z}_i$.  
Each link must interact with four ancillary pseudospins in the limit $T\ll \chi$ or $\chi\ll g$ to allow thermalization to the ground state or the thermal state of $H^{TC}_0$, respectively. 

\begin{figure}[t] 
   \centering
  \includegraphics[width=3.3in]{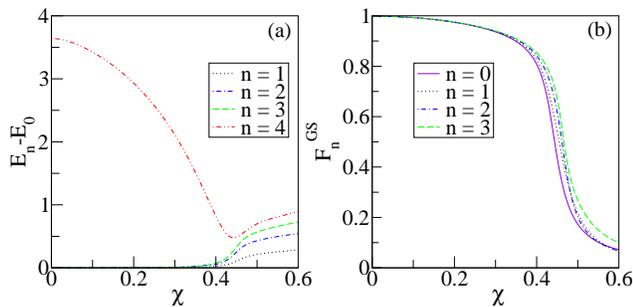} 
 \caption{  (a) Energy spectrum (in units of $J_e = J_m$) and (b) ground state fidelity vs. perturbation strength from exact diagonalization of the 18 site toric code with $h_z = 0.05$.}
\label{fig3}
\end{figure}

The stroboscopic generation of $H_0^{TC}$ outlined above introduces truncation perturbations in the perturbative expansion, e.g, the second term in Eq.~(\ref{Heffzzzz}),
 which are distinct from extrinsic errors due to experimental noise and gate inaccuracies.  If sufficiently large, such truncation perturbations could drive the system away from the desired ground state phase
. We now show on a finite sized system accessible to current experiments~\cite{nelson2007}, that the intrinsic perturbations can be kept sufficiently small.
Fig.~\ref{fig3}(a) plots the gap of $H^{TC} _{\chi,h_z} = H_0^{TC}-h_z \sum_i \sigma^z_i+\chi \sum_{\langle i,j \rangle} \sigma_i^x\sigma_j^y$ as a function of the strength of the perturbation for a 3$\times$3 planar lattice with toroidal boundary conditions (Fig.~\ref{fig2})
. The Zeeman field is added here to fully split the ground state degeneracy and ensure robust characterization of the eigenstates of $H$ even in the presence of small additional perturbations.
We define the ground state fidelity as $F^{GS}_n = \vert \langle \Psi^0_n \vert  \Psi_n \left(\chi,h_z \right) \rangle \vert$, where the $\left \vert  \Psi^0_n \right \rangle$ are the degenerate ground states of $H^{TC}_0$ and $\left \vert  \Psi_n \left(\chi,h_z \right) \right \rangle$ are the nearly degenerate ground states of $H^{TC} _{\chi,h_z}$. Fig.~\ref{fig3}(b) shows the ground state fidelity as a function of $\chi$. This fidelity determines the robustness of topological operations that will be performed via string operators~\cite{kitaev2003} to measure or perform
gates on the system. We see that for $\left \vert \chi \right \vert \lesssim 0.4$ the features of the topological phase persist, including the approximate four-fold degeneracy of the ground state and a finite gap to excitations. This corresponds to a maximum value of $\phi \sim 0.4$, which constrains the gate operations in the pulse sequences, Eq.~(\ref{eq:sequence}). This robustness should increase with increasing lattice size, and is consistent with known stability of 
$H^{TC}_0$
to perturbations \cite{trebst2007}.

Increasing $\phi$ increases $J$ and therefore the gap of the $H^{TC}_0$; however it also increases $\chi/J$, which reduce the gap of $H^{TC} _{\chi,h_z}$ and topological protection for large $\chi$. We also note that $J \sim 1/N_G$ where $N_G$ is the number of sequential gates used to simulate $H$. 
For larger lattices, some degree of parallelization is thus desirable to ensure that the gap does not decrease with the lattice size. Choosing $\chi=0.2$, the  gap  achieved by a completely serial implementation is $\Delta \approx 0.6~\mu\mathrm{K}/N_{\mathrm{sys}}$, where $N_\mathrm{sys}$ is the number of system atoms used. With the cooling sequence serially interleaved, $\Delta \approx 0.1~\mu\mathrm{K}/N_{\mathrm{sys}}$ and $\lambda^* \sim 10^{4}~\mathrm{s}^{-1}/N_{\mathrm{sys}}$ are achievable~\cite{longpaper}. For a minimal system of 18 system atoms, this allows for an effective temperature $T_{\mathrm{eff}} < \Delta$ to be reached with an error rate of $\mathrm{EPG} \sim 10^{-4}$ or less~\cite{longpaper}.

\prsec{Sources of Errors}   
This scheme is designed to be robust against errors within the pseudospin subspace.  
The dominant source of residual error in the implementation discussed here is leakage from the Rydberg levels due to spontaneous emission and black body radiation.  The latter may be effectively suppressed by working at low temperatures~\cite{hessels1998}, and spontaneous emission is minimized by utilizing states with larger $n$. 
With $n\lesssim180$, we estimate that spointaneous emission errors can be reduced to $\sim\!\!10^{-6}$ per gate, allowing for up to $10^3$ stroboscopic cycles. 

\prsec{Discussion}We have developed a formalism 
for the stroboscopic generation of $n$-body Hamiltonians using $1$- and $2$-body quantum operations together with a dissipative thermalization scheme.   
We have applied this 
here to the toric code Hamiltonian in the context of  
addressable optical lattice experiments \cite{nelson2007}. 
Our method applies to a wide range of lattice spin models ~\cite{longpaper} as well as to other experimental setups~\cite{Porto2003}.  
The dynamic generation both of a Hamiltonian possessing a topologically ordered ground state and of an effective thermalization mechanism offers the possibility of \emph{robust} simulation of the ground state and of the creation and braiding of anyonic excitations~\cite{han2007}.
These are essential components required for the topologically protected storage and manipulation of quantum information.

We thank D. Weiss for useful discussions. This material is based upon work supported by DARPA under Award No. 3854-UCB-AFOSR-0041. During the preparation of this manuscript related results discussing ground state preparation were reported \cite{Weimer2009}.

\bibliography{bibliography}
\bibliographystyle{apsrev}

\end{document}